\documentclass[%
reprint,
superscriptaddress,
 amsmath,amssymb,
 aps,
]{revtex4-2}

\usepackage{graphicx}
\usepackage{dcolumn}
\usepackage{bm}
\usepackage{relsize}
\usepackage[usenames,dvipsnames]{xcolor}
\usepackage{hyperref}

\begin{document}

\title{Mixed-valence state in the dilute-impurity regime of La-substituted SmB$_6$}

\vspace{5mm}

\author{M. Zonno}
\email[]{marta.zonno@synchrotron-soleil.fr}
\thanks{\\ Current address: Synchrotron SOLEIL, Saint-Aubin 91192, France}
\affiliation{Department of Physics \& Astronomy, University of British Columbia, Vancouver, BC V6T 1Z1, Canada}
\affiliation{Quantum Matter Institute, University of British Columbia, Vancouver, BC V6T 1Z1, Canada}
\affiliation{Canadian Light Source Inc., Saskatoon, SK S7N 2V3, Canada}
\author{M. Michiardi}
\affiliation{Department of Physics \& Astronomy, University of British Columbia, Vancouver, BC V6T 1Z1, Canada}
\affiliation{Quantum Matter Institute, University of British Columbia, Vancouver, BC V6T 1Z1, Canada}
\affiliation{Max Planck Institute for Chemical Physics of Solids, Dresden 01187, Germany}
\author{F. Boschini}
\affiliation{Quantum Matter Institute, University of British Columbia, Vancouver, BC V6T 1Z1, Canada}
\affiliation{Centre \'{E}nergie Mat\'{e}riaux T\'{e}l\'{e}communications Institut National de la Recherche Scientifique, Varennes, QC J3X 1S2, Canada}
\author{G. Levy}
\affiliation{Department of Physics \& Astronomy, University of British Columbia, Vancouver, BC V6T 1Z1, Canada}
\affiliation{Quantum Matter Institute, University of British Columbia, Vancouver, BC V6T 1Z1, Canada}
\author{K. Volckaert}
\author{D. Curcio}
\author{M. Bianchi}
\affiliation{Department of Physics and Astronomy, Interdisciplinary Nanoscience  Center, Aarhus University, 8000 Aarhus C, Denmark}
\author{\linebreak P. F. S. Rosa}
\affiliation{Los Alamos National Laboratory, Los Alamos, New Mexico 87545, USA}
\author{Z. Fisk}
\affiliation{Department of Physics and Astronomy, University of California, Irvine, California 92697, USA}
\author{Ph. Hofmann}
\affiliation{Department of Physics and Astronomy, Interdisciplinary Nanoscience  Center, Aarhus University, 8000 Aarhus C, Denmark}
\author{I. S. Elfimov}
\affiliation{Department of Physics \& Astronomy, University of British Columbia, Vancouver, BC V6T 1Z1, Canada}
\affiliation{Quantum Matter Institute, University of British Columbia, Vancouver, BC V6T 1Z1, Canada}
\author{R. J. Green}
\affiliation{Quantum Matter Institute, University of British Columbia, Vancouver, BC V6T 1Z1, Canada}
\affiliation{Department of Physics \& Engineering Physics, University of Saskatchewan, Saskatoon, SK S7N 5E2, Canada}
\author{G. A. Sawatzky}
\affiliation{Department of Physics \& Astronomy, University of British Columbia, Vancouver, BC V6T 1Z1, Canada}
\affiliation{Quantum Matter Institute, University of British Columbia, Vancouver, BC V6T 1Z1, Canada}
\author{A. Damascelli}
\email[]{damascelli@physics.ubc.ca}
\affiliation{Department of Physics \& Astronomy, University of British Columbia, Vancouver, BC V6T 1Z1, Canada}
\affiliation{Quantum Matter Institute, University of British Columbia, Vancouver, BC V6T 1Z1, Canada}

\begin{abstract}
\noindent
Homogeneous mixed-valence (MV) behaviour is one of the most intriguing phenomena of $f$-electron systems. Despite extensive efforts, a fundamental aspect which remains unsettled is the experimental determination of the limiting cases for which MV emerges. Here we address this question for SmB$_6$, a prototypical MV system characterized by two nearly-degenerate Sm$^{2+}$ and Sm$^{3+}$ configurations. By combining angle-resolved photoemission spectroscopy (ARPES) and x-ray absorption spectroscopy (XAS), we track the evolution of the mean Sm valence, $v_{Sm}$, in the Sm$_x$La$_{1-x}$B$_6$ series. Upon substitution of Sm ions with trivalent La, we observe a linear decrease of valence fluctuations to an almost complete suppression at $x$\,=\,0.2, with $v_{Sm}$\,$\sim$\,2; surprisingly, by further reducing $x$, a re-entrant increase of $v_{Sm}$ develops, approaching the value of $v_{imp}$\,$\sim$\,2.35 in the dilute-impurity limit. Such behaviour departs from a monotonic evolution of $v_{Sm}$ across the whole series, as well as from the expectation of its convergence to an integer value for $x$\,$\rightarrow$\,0. 
Our ARPES and XAS results, complemented by a phenomenological model, demonstrate an unconventional evolution of the MV character in the Sm$_x$La$_{1-x}$B$_6$ series, paving the way to further theoretical and experimental considerations on the concept of MV itself, and its influence on the macroscopic properties of rare-earth compounds in the dilute-to-intermediate impurity regime.
\end{abstract}

\maketitle

\section*{Introduction}
\noindent
Strong many-body interactions play a critical role in shaping the electronic, magnetic, and even mechanical properties of quantum materials. In compounds containing rare-earth or actinide elements, electron correlations originate from the localized and partially filled $f$-electron shells. The resulting entanglement of the relevant degrees of freedom -- orbital, spin, charge, and lattice -- gives rise to a plethora of novel phenomena in such materials, superconductivity \cite{mathur1998magnetically, moriya2003antiferromagnetic}, including spin and charge order \cite{hotta2006orbital}, quantum criticality \cite{gegenwart2008quantum}, heavy fermion behavior \cite{fisk1988heavy, coleman2015heavy}, Kondo physics \cite{kondo1964resistance}, and mixed-valence behaviour \cite{varma1976mixed, peter2007molecules}. 

In particular, mixed valence (MV) is a fascinating phenomenon observed in a wide range of rare-earth compounds \cite{varma1976mixed, lawrence1981valence, buschow1979intermetallic, coey1999mixed, parks2012valence}, yet a full microscopic understanding of its nature and limits remains elusive. MV is defined by the presence of a given rare-earth element in the system exhibiting more than one electronic occupation for the $f$ shell \cite{Anderson1984}. Within the whole class of MV compounds, an important distinction arises between inhomogeneous and homogeneous MV scenarios. While in the former case ions with differing $f$ occupation values reside on inequivalent crystallographic sites, in the latter all rare-earth ions retain the same non-integer $f$ valence at each site \cite{Riseborough2016, Riseborough2000}.
Here, we focus on the case of homogeneous MV -- also referred to as intermediate valence in the literature \cite{Riseborough2016} -- to explore experimentally the parameter space in which such MV behaviour emerges. Various theoretical works have discussed the phenomenon of MV in the dilute- to single-impurity limit \cite{BickersRMP, QimiaoSi1996}, from mean-field theories \cite{haldane1977new, VarmaHeinePRB} to exact solutions by Wilson's renormalization group method or bosonization \cite{perakis1993model, sire1994model}.
However, an experimental study tracking the crossover of the MV character going from a periodic $f$-electron lattice to a dilute $f$-impurity system is still lacking.

\begin{figure*}
\includegraphics[scale=1]{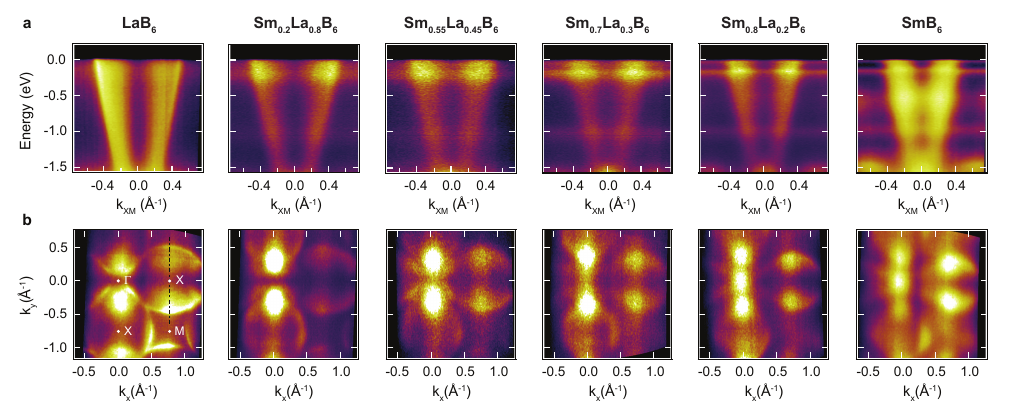}
\caption{\textbf{ARPES spectra of Sm$_x$La$_{1-x}$B$_6$.} \textbf{a} ARPES spectra along the $\overline{\text{XM}}$ high-symmetry direction of the Brillouin zone (black dashed line in \textbf{b}, left) for $x$\,=\,[0, 0.2, 0.55, 0.7, 0.8, 1]. \textbf{b} ARPES iso-energy contours close to E$_F$ for the same samples shown in \textbf{a}; the integration window in energy is 15\,meV about E$_F$. All data were acquired at 10\,K with $h\nu$\,=\,21.2\,eV.}
\label{fig:Spectra} 
\end{figure*}

In this work, we address this question by focusing on the prototypical homogeneous mixed-valence system SmB$_6$, wherein the interplay between two nearly-degenerate $f$-shell valence configurations of the Sm ions profoundly shapes its electronic structure and macroscopic properties.
While it has been shown that temperature and pressure may be exploited as external perturbations to tune the intermediate valence of the Sm ions \citep{tarascon1980temperature, mizumaki2009temperature, butch2016pressure, sun2016pressure, lee2017fluctuating, Zhou2017pressure, emi2018temperature}, we base our experimental strategy on elemental substitution on the rare-earth site. This approach provides a powerful chemical control parameter acting directly on the occupation of the $f$-states, allowing the precise tracking of the mean Sm valence across different concentration regimes.

To this end, we employ trivalent La ions as substituent in the $\mathrm{Sm}_x\mathrm{La}_{1-x}\mathrm{B}_6$ hexaboride series. Although all the compounds of the series share the same CsCl-type crystal structure, the two end members exhibit very different physics. LaB$_6$ ($x$\,=\,0) is metallic owing to the partially occupied La-$5d$ band and the lack of 4$f$ electrons. In contrast, the precise nature of the ground state of $\mathrm{SmB}_6$ ($x$\,=\,1) still remains an open question. 
Exhibiting a resistivity plateau at low temperature \cite{menth1969magnetic}, it has been theoretically proposed as realization of a topological Kondo insulator \cite{dzero2010topological, dzero2012theory}, and various experimental studies have later discussed the possible presence and nature of in-gap electronic states \cite{wolgast2013low, zhang2013hybridization, kim2013surface, jiang2013observation, xu2013surface, neupane2013surface, frantzeskakis2013kondo, zhu2013polarity, hlawenka2018samarium, varma2020majoranas}, as well as controversial reports of quantum oscillations \cite{li2014two, tan2015unconventional, Thomas2019}.
Even though a clear answer has yet to emerge, a fundamental aspect characterizing the physics of $\mathrm{SmB}_6$ is 
undoubtedly the nearly-complete admixture of the two possible Sm ions valence configurations $4f^6\,5d^0$ and $4f^5\,5d^1$, which in terms of the $f$-level occupation are generally referred to as Sm$^{2+}$ and Sm$^{3+}$, respectively. This leads to a mean Sm valence for the $f$ shell of $+$\,2.505 in SmB$_6$ at low temperature \cite{mizumaki2009temperature, tarascon1980temperature, sundermann4fcrystalfield}, while the single $d$ band is characterized by a strongly mixed B-Sm character.

A recent de Haas–van Alphen (dHvA) investigation of dilute Sm-doped LaB$_6$ ($x$=0.05 and $x$=0.1) reported only a small reduction of the FS volume upon Sm substitution \cite{Goodrich2009dHvA}. Interestingly, such reduction rate would not be compatible with having Sm ions either purely divalent or trivalent in this concentration regime.
Here, we track the electronic structure of the $\mathrm{Sm}_x\mathrm{La}_{1-x}\mathrm{B}_6$ series over the entire doping range, by means of angle-resolved photoemission (ARPES) and x-ray absorption spectroscopy (XAS), and observe a non-monotonic evolution of the mean Sm valence. While the strong Sm$^{2+}$/Sm$^{3+}$ admixture is quenched in the intermediate substitution regime, it resurges for low Sm concentrations, with a persisting MV behavior all the way into the dilute-impurity limit. These results provide experimental evidence of the emergence of the MV phenomenon even in this dilute limit, and establish the key role of unconventional behaviour of $f$-electrons in defining the properties of rare-earth compounds also in such extreme regime.

\begin{figure*}
\includegraphics[scale=1]{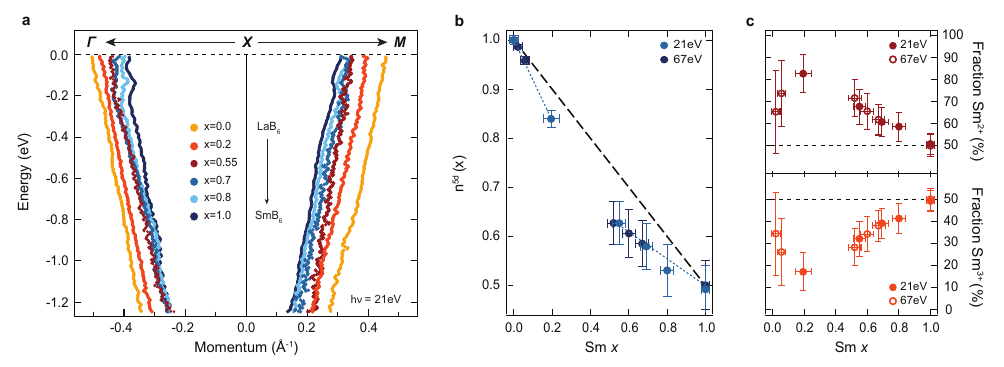}
\caption{\textbf{Extracting the Sm valence from the ARPES dispersion.} \textbf{a} Evolution of the X-pocket dispersion along $\overline{\Gamma \text{X}}$ and $\overline{\text{XM}}$ directions for $x$\,=\,[0, 0.2, 0.55, 0.7, 0.8, 1], as extracted from the ARPES spectra in Fig.\,\ref{fig:Spectra}. \textbf{b} Electronic occupation number of the X-pocket, $n^{5d} (x)$, for different $\mathrm{Sm}_x \mathrm{La}_{1-x} \mathrm{B}_6$ compounds. Data points were obtained from ARPES spectra acquired with $21.2\,\mathrm{eV}$ and $67\,\mathrm{eV}$. The dashed black line illustrates the case of a constant 1:1 ratio of $\mathrm{Sm}^{2+}$:\,$\mathrm{Sm}^{3+}$ across the series; this is computed from the general expression $n^{5d}$\,=\,$[1-x*a/(a+b)]$, where $a$ and $b$ are the fractional percentage of $\mathrm{Sm}^{2+}$ and \,$\mathrm{Sm}^{3+}$, which reduces to $n^{5d}_{1:1}$\,=\,$1-x/2$ in the case of $a=b$. \textbf{c} Calculated fractional percentage of $\mathrm{Sm}^{2+}$ (top) and $\mathrm{Sm}^{3+}$ (bottom) by using Eq.\,\ref{eq:eq1} for the different compounds measured by ARPES. Error bars in \textbf{b} and \textbf{c} are determined as follow: on the $x$ axis are based on energy dispersive x-ray (EDX) measurements; on $n^{5d} (x)$ are derived from the fitting of the ARPES data; on the fractions of $\mathrm{Sm}^{2+}$/$\mathrm{Sm}^{3+}$ are calculated by combining the uncertainties on $x$ and $n^{5d} (x)$ via error propagation rules.}
\label{fig:ARPES} 
\end{figure*}

\section*{Results}
We begin by showcasing the evolution of the electronic structure of $\mathrm{Sm}_x \mathrm{La}_{1-x} \mathrm{B}_6$, upon La-Sm chemical substitution, as measured by ARPES. Figure\,\ref{fig:Spectra}a summarizes the ARPES spectra acquired along the $\overline{\text{XM}}$ direction (black dashed line in Fig.\,\ref{fig:Spectra}b, left) for $x$\,=\,[0, 0.2, 0.55, 0.7, 0.8, 1]. 
Common to all compounds, bulk electron-like pockets are centered at the X high-symmetry points of the Brillouin zone (BZ), forming elliptic iso-energy contours (see Fig.\ref{fig:Spectra}b). Being primarily associated with $\mathrm{B}$-$2p$ and rare-earth $5d$ electrons, their size is directly related to the valence of the rare-earth element in the material: for a full 3+ configuration the $d$-pocket is half filled and at its largest, while in a 2+ state the valence electrons only fall in the $f$-states and the $d$-pocket lays in the unoccupied part of the spectrum. This observation makes the study of the evolution of the bulk $5d$ pockets centered at X instrumental to track the possible valence fluctuations of the Sm ions in the $\mathrm{Sm}_x \mathrm{La}_{1-x} \mathrm{B}_6$ series.

Note that as Sm is introduced into the system, three non-dispersing $4f$-states emerge in the ARPES spectra within the first 1\,eV of the Fermi level. Starting very weak and broad for small $x$ values, these states become gradually sharper and shift to slightly lower binding energies as $x$ increases, finally settling at 15\,meV, 150\,meV, and 1\,eV for pristine $\mathrm{Sm}\mathrm{B}_6$, consistent with the values reported in the literature \citep{jiang2013observation, neupane2013surface, denlinger2014smb6}. Concurrently, the size of the X-pockets progressively decreases upon increasing $x$. While the interplay between localized $4f$-electrons and itinerant $5d$-electrons is undoubtedly an important defining aspect of the electronic structure of these materials, in this work we focus our analysis on the evolution of the bulk X-pockets' dispersion as a function of $x$, which can be directly linked to changes in the concentration of trivalent ions in the system.  

While a decrease in the size of the X-pockets is expected throughout the series due to the removal of trivalent La ions contributing to the occupation of the $5d$-band, by visual inspection of Fig.\ref{fig:Spectra} we note that the observed behaviour departs from a constant-rate reduction. In fact, the change in the ARPES dispersion is more pronounced for $x\leq0.55$ than for higher Sm concentrations. This progression is showcased by the evolution of the $5d$-band dispersion extracted as a function of $x$ along the $\overline{\Gamma \text{X}}$ and $\overline{\text{XM}}$ high-symmetry directions shown in Fig.\,\ref{fig:ARPES}a. In order to quantitatively assess this variation and thus establish a direct relation between the ARPES dispersion and the fractional percentage of $\mathrm{Sm}^{2+}$ and $\mathrm{Sm}^{3+}$ present in the system, we must convert the size of the X-pocket contours as extracted from the ARPES data into the pocket's occupation $n^{5d}$. This is done via application of the Luttinger's theorem, which directly relates the volume enclosed by a material's Fermi surface to the electron density \citep{martin1982fermi, martin1982fermi2}.
Here we emphasize that the bare X-pocket dispersion in the $\mathrm{Sm}_x \mathrm{La}_{1-x} \mathrm{B}_6$ series can be described to a first approximation by the same effective mass as observed for $\mathrm{La}\mathrm{B}_6$, with the only Fermi momentum $k_F^{5d}$ changing to accommodate for the varying electronic occupation. This assumption is supported by the ARPES dispersions shown in Fig.\,\ref{fig:ARPES}a, and facilitates a direct comparison among different compounds in the series.

\begin{figure*}[t]
\includegraphics[scale=1]{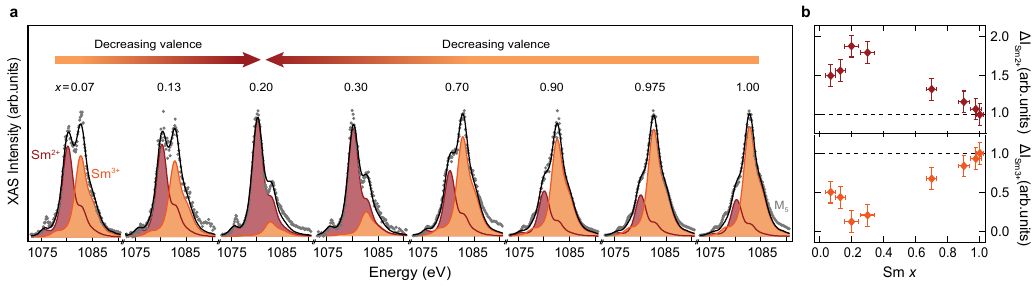}
\caption{\textbf{XAS study of the Sm$_x$La$_{1-x}$B$_6$ series.} \textbf{a} Evolution of the XAS intensity at the Sm $M_5$ edge for $x$\,=\,[0.07, 0.13, 0.2, 0.3, 0.7, 0.9, 0.975, 1]. The absorption profiles (grey dots) have been extracted at 640\,eV (La fluorescence line) of IPFY spectra for $x$\,$\leq$\,0.3, and at 850\,eV (Sm fluorescence line) of PFY spectra for $x$\,$>$\,0.3. The total XAS spectral weight is fit (black lines) by the sum of two independent components associated to $\mathrm{Sm}^{2+}$ (red lines and shaded regions) and $\mathrm{Sm}^{3+}$ ions (orange lines and shaded regions). \textbf{b} Intensity evolution of the $\mathrm{Sm}^{2+}$ (top) and $\mathrm{Sm}^{3+}$ (bottom) components normalized to the $x$\,=\,1 case. All data were taken at a base temperature of $20\,\mathrm{K}$. Error bars in \textbf{b} are determined as follow: on the $x$ axis are based on EDX measurements; on $\Delta$I$_{\mathrm{Sm}^{2+}/\mathrm{Sm}^{3+}}$ are derived from the fitting of the XAS data.}
\label{fig:XAS} 
\end{figure*}

Figure\,\ref{fig:ARPES}b displays the values of $n^{5d}(x)$ as extracted via Luttinger's theorem from the ARPES spectra acquired with $21.2 \, \mathrm{eV}$ and $67 \, \mathrm{eV}$ probe energy (the latter associated to the bulk $\mathrm{\Gamma}$ high-symmetry point; see Supplementary Information). 
The values were obtained by calculating the total enclosed volume of the X-pockets based on the collected ARPES while relying on the cubic crystal sysmmetry, and by taking into account the reported variation of the lattice parameter in the $\mathrm{Sm}_x \mathrm{La}_{1-x} \mathrm{B}_6$ series \cite{kasaya1980study, tarascon1980valence}.
For $\mathrm{La}\mathrm{B}_6$ ($x$\,=\,0), the pockets enclose $\sim$\,50\,$\%$ of the bulk cubic BZ, corresponding to having 1 electron $n^{5d}$\,=\,1. At the other end of the series, in $\mathrm{Sm}\mathrm{B}_6$ ($x$\,=\,1) only a quarter of the cubic BZ is filled by the X-pockets, yielding $n^{5d}$\,=\,0.5.
These results are fully consistent with the metallic ground state observed in $\mathrm{La}\mathrm{B}_6$ and the reported valence of +2.505 of Sm ions in $\mathrm{Sm}\mathrm{B}_6$, thus validating our analysis. By applying the same approach to the intermediate compounds of the series, we find that $n^{5d}(x)$ clearly deviates from the linear reduction expected in the case of a constant 1:1 ratio of $\mathrm{Sm}^{2+}$:\,$\mathrm{Sm}^{3+}$, represented in Fig.\,\ref{fig:ARPES}b by the black dashed line. To better quantify the evolution, we compute the fractional percentage of $\mathrm{Sm}^{2+}$ and $\mathrm{Sm}^{3+}$ from $n^{5d}(x)$, as follows (normalized over the total amount of Sm in the system, $x$):
\begin{equation}
    \begin{split}
        \mathrm{Fraction}\,\,\mathrm{Sm}^{2+} \, \mathrm{(\%)} & = \frac{1-n^{5d}(x)}{x}\\
        \mathrm{Fraction}\,\,\mathrm{Sm}^{3+} \, \mathrm{(\%)} & = 1+\frac{n^{5d}(x)-1}{x} \quad .
    \end{split}
    \label{eq:eq1}
\end{equation}
The resulting values are presented in Fig.\,\ref{fig:ARPES}c. 
As $x$ decreases from 1, the amount of $\mathrm{Sm}^{2+}$ gradually increases upon reaching a maximum of $\sim$\,85\,$\%$ at $x$\,=\,0.2, followed by a re-entrant reduction at even lower Sm concentrations. 
Despite the uncertainties associated with probing minimal modifications of the X-pocket ARPES dispersion for concentrations smaller than 0.1 (as reflected in the large error bars in Fig.\,\ref{fig:ARPES}c), our ARPES results suggest a clear distinction between the low ($x\leq0.2$) and high ($x\geq0.55$) Sm concentration regimes, along with a substantial variation of the $\mathrm{Sm}^{2+}$:\,$\mathrm{Sm}^{3+}$ ratio across the $\mathrm{Sm}_x \mathrm{La}_{1-x} \mathrm{B}_6$ series.

In an effort to verify this scenario and gain more insights on the low concentration regime also in connection to the previous dHvA work on dilute Sm-doped LaB$_6$ \cite{Goodrich2009dHvA}, we complemented the ARPES data with a XAS study of the same $\mathrm{Sm}_x \mathrm{La}_{1-x} \mathrm{B}_6$ series. In particular, to achieve a higher bulk sensitivity and thus establish our results as an intrinsic bulk-property, we exploited partial and inverse partial fluorescence yield (PFY and IPFY). Both of these technique are characterized by a probing depth of tens of nm (thus excluding significant surface-related contributions to the XAS signal; see Supplementary Information for details), allowing one to circumvent some of the challenges characteristic of ARPES on $\mathrm{Sm}_x \mathrm{La}_{1-x} \mathrm{B}_6$, such as cleaving and surface degradation. Furthermore, XAS has already been shown to be a powerful technique to explore the physics of MV systems, such as $\mathrm{SmB}_6$: the absorption spectrum can be described to first approximation by the sum of two independent components, corresponding to $\mathrm{Sm}^{2+}$ and $\mathrm{Sm}^{3+}$ ions. By tuning the incident energy across the Sm $M_4$ and $M_5$ edges (i.e. exciting $3d$ core electrons into $4f$ orbitals), each XAS spectrum can be mapped into a specific $\mathrm{Sm}^{2+}$:\,$\mathrm{Sm}^{3+}$ ratio, providing us with a tool to directly determine the mean Sm valence in the $\mathrm{Sm}_x \mathrm{La}_{1-x} \mathrm{B}_6$ series. 

In Fig.\,\ref{fig:XAS}a we present the evolution of the XAS intensity at the Sm $M_5$ edge for $x$\,=\,[0.07, 0.13, 0.2, 0.3, 0.7, 0.9, 0.975, 1], along with a weighted sum fit of the $\mathrm{Sm}^{2+}$ and $\mathrm{Sm}^{3+}$ components (red and orange, respectively). While for high $x$ the $\mathrm{Sm}^{3+}$ component dominates, its contribution dramatically reduces at $x$\,=\,0.2. However, upon further decrease of $x$, a clear inversion in the progression of the XAS spectra is observed, as the $\mathrm{Sm}^{3+}$ component strengthens again for $x\leq0.2$. 

Such behaviour is highlighted by the intensity variation of the two components displayed in Fig.\,\ref{fig:XAS}b (normalized to the $x$\,=\,1 values) and is fully consistent with the evolution of the fractional percentages of $\mathrm{Sm}^{2+}$ and $\mathrm{Sm}^{3+}$ obtained from ARPES, thus confirming the two-regime scenario already suggested in Fig.\,\ref{fig:ARPES}. 
In particular, the $\mathrm{Sm}^{2+}$ contribution peaks nearly doubling at $x$\,=\,0.2, corresponding to an increment of $\mathrm{Sm}^{2+}$ ions in the system as large as $\sim 90\%$ with respect to the pure SmB$_6$ case.
We remark that such significant increase of $\mathrm{Sm}^{2+}$ sets apart from what reported on pure $\mathrm{Sm}\mathrm{B}_6$ by employing high temperature and high pressure, with both perturbations causing an increase of the mean Sm valence towards +3 \citep{tarascon1980temperature, mizumaki2009temperature, butch2016pressure, sun2016pressure, lee2017fluctuating, Zhou2017pressure, emi2018temperature}.
Also note that the deviation from an almost 1:1 ratio of the $\mathrm{Sm}^{2+}$:$\mathrm{Sm}^{3+}$ peaks expected at $x$\,=\,1 may reflect a difference in the relative cross-sections of the two Sm components at the energies the XAS measurements were performed: while not affecting the qualitative evolution of the intensities shown in Fig.\,\ref{fig:XAS}b, it is taken into account for computing the mean Sm valence (see Supplementary Information for details on the XAS analysis and normalization).

\begin{figure}
\includegraphics[scale=1]{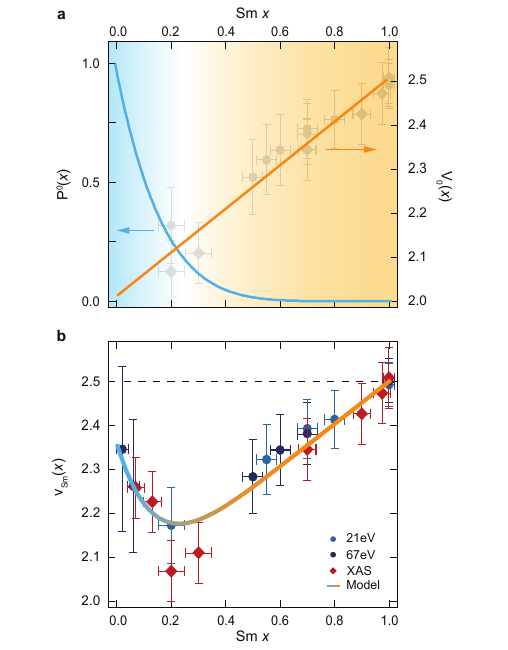}
\caption{\textbf{Simplified model for the description of $v_{Sm}$ in the two different $x$ regimes.} \textbf{a} Light blue line: calculated probability for a Sm ions to have no other Sm as next-nearest-neighbours, $P^0 (x)$. This function is used to define the fraction of Sm ions which acquires the impurity valence $v_{imp}$\,=\,2.35 at each $x$. Orange line: linear fit to the experimental $v_{Sm}$ values for $x$\,$\geq$\,0.2 (grey data points), $V_0(x)$, capturing the decrease in the valence due to formation of $\mathrm{Sm}^{2+}$ upon La-substitution. Light blue and yellow shadowed regions indicate the Sm concentrations regimes at which $P^0 (x)$ and $V_0(x)$ dominates, respectively. \textbf{b} Comparison between the model of Eq.\ref{eq:eq2} (solid line; light blue/yellow color indicates the two different contributions to our model shown in \textbf{a})  and the experimental values of $v_{Sm}$ obtained from ARPES (light and dark blue circles, acquired with $21$\,eV and $67$\,eV, respectively) and XAS (red diamonds). Error bars in \textbf{b} are determined as follow: on the $x$ axis are based on EDX measurements; on $v_{Sm}$ are derived from the uncertainties on the fractions of $\mathrm{Sm}^{2+}$/$\mathrm{Sm}^{3+}$ and on $\Delta$I$_{\mathrm{Sm}^{2+}/\mathrm{Sm}^{3+}}$ for ARPES and XAS data, respectively.}
\label{fig:Model} 
\end{figure}

In Fig.\ref{fig:ARPES} and \ref{fig:XAS} we showed that both ARPES and XAS measurements of the $\mathrm{Sm}_x \mathrm{La}_{1-x} \mathrm{B}_6$ series display a progression from an evenly $\mathrm{Sm}^{2+}$/$\mathrm{Sm}^{3+}$ regime into a predominant presence of $\mathrm{Sm}^{2+}$ as $x$ decreases. This result is consistent with the observation of the average Sm valence tending towards +2 upon trivalent ion substitution (such as $\mathrm{La}^{3+}$ or $\mathrm{Y}^{3+}$) reported in early works \cite{kasaya1980study, tarascon1980valence}. Furthermore, recent transport studies on La-substituted $\mathrm{Sm}\mathrm{B}_6$ have reported the complete closure of the $d$-$f$ hybridization gap, and the consequent emergence of a metallic-like behaviour, for La concentrations higher than $25\%$ (here $x\leq0.75$) \cite{kang2016magnetic, gabani2016transport}, corroborating the substantial increase of $\mathrm{Sm}^{2+}$ in the system observed in this work. 
However, according to these arguments one may expect the valence fluctuations to be quenched at zero doping, i.e. $v_{Sm}$\,$\rightarrow$\,+2 for $x$\,$\rightarrow$\,0, in stark contrast with the clear suppression detected at $x$\,=\,0.2, followed by the sudden overturn for even lower $x$.
Nevertheless, a scenario in which the Sm ions are neither purely divalent nor trivalent even in the dilute-impurity regime is in agreement with the evolution of the FS volume reported by dHvA, which exhibits a reduction rate smaller than the amount of Sm introduced in the system \cite{Goodrich2009dHvA}.

Here we present a basic phenomenological model for $v_{Sm}$ in the $\mathrm{Sm}_x \mathrm{La}_{1-x} \mathrm{B}_6$ series based on the two distinct regimes observed experimentally. On one hand, for high $x$ we mimic the convergence towards +2 upon La substitution (i.e., upon reducing $x$) by fitting the experimental data for $0.2\leq x \leq 1$ with a linear fit, $V_0(x)$ [orange line in Fig.\ref{fig:Model}a]. 
On the other hand, to provide a phenomenological description of the increase of $v_{Sm}$ detected at low $x$ (i.e., $x<0.2$), 
in connection to the literature discussing the MV behaviour in the extreme dilute limit of a single magnetic impurity in a non-magnetic band metal, we consider here the possibility of additional contributions to $v_{Sm}$ stemming from Sm ions acting as single impurities with a specific fixed fractional valence, $v_{imp}$.
In this regard, we compute the probability as a function of $x$ for a Sm ion to have no other Sm in the next-nearest-neighbour sites, $P^0 (x)$ [light blue line in Fig.\ref{fig:Model}a]. This function is used to define in first approximation the fraction of Sm ions that exhibits a dilute-impurity valence $v_{imp}$\,=\,+2.35, as extrapolated from the experimental results in Figs.\,\ref{fig:ARPES}c\,-\,\ref{fig:XAS}b, at any given concentration $x$. 
We can then express the evolution of the Sm valence in the $\mathrm{Sm}_x \mathrm{La}_{1-x} \mathrm{B}_6$ series as:
\begin{equation}
 \mathrm{v}_{model}=P^0(x) \cdot v_{imp} + [1-P^0(x)] \cdot V_0(x) \quad.
 \label{eq:eq2}
\end{equation}
Fig.\ref{fig:Model}b compares the model of Eq.\,\ref{eq:eq2} with the experimental values of the mean Sm valence obtained from ARPES (light and dark blue circles) and XAS (red diamonds), showing an overall good agreement. In particular, the inclusion of the emerging Sm dilute-impurity regime with fixed fractional valence is proven pivotal to capture the steep increase of $v_{Sm}$ experimentally observed for low $x$; however, its contribution becomes negligible at high $x$, owing to the rapid decay of $P^0 (x)$ below 0.1 for $x$\,$\geq$\,0.3. 

As a final note, we emphasize that the model of Eq.\,\ref{eq:eq2} does not fully describe the nearly-complete suppression of $\mathrm{Sm}^{2+}$/$\mathrm{Sm}^{3+}$ admixture detected at $x$\,=\,0.2. 
When performing density functional theory calculations of the doping dependence on the Fermi energy using virtual crystal approximation, no anomalous behaviour was found which could explain this observation; indeed additional investigations are needed to specifically address the sharp crossover observed around $x$\,=\,0.2 with the development of more refined theories. 
Nevertheless, our combined ARPES and XAS study provides evidence of the realization of a dilute-impurity MV state in the $\mathrm{Sm}_x \mathrm{La}_{1-x} \mathrm{B}_6$ series. Our results may stimulate further theoretical and experimental considerations on the concept of mixed valence and its influence on the macroscopic electronic and transport properties of rare-earth compounds in the dilute-to-intermediate impurity regime.

\section*{\label{}Methods}
High-quality single crystals of $\mathrm{Sm}_x \mathrm{La}_{1-x} \mathrm{B}_6$ were grown by the aluminum flux method in a continuous Ar-purged vertical high-temperature tube furnace \cite{wolgast2013low}. Post-growth characterization by scanning electron microscope and energy dispersive x-ray measurements for the actual Sm concentration was performed at the Center for Integrated Nanotechnologies, an Office of Science User Facility operated for the U.S. Department of Energy Office of Science. ARPES experiments were performed at the Stewart Blusson Quantum Matter Institute at UBC employing a photon energy of $h\nu$\,=\,21.2\,eV, at a base pressure $<$\,3\,$\cdot 10^{-11} $\,Torr and base temperature of 10\,K. The electrons were collected using a SPECS Phoibos 150 hemisperical analyzer, with energy and momentum resolution of 25\,meV and 0.02\,$\mathrm{\AA}$, respectively. Additional ARPES measurements were carried out at the SGM3 endstation at the ASTRID2 synchrotron radiation facility \cite{hoffmann2004undulator}, using a photon energy of $h\nu$\,=\,67\,eV, with base temperature 35\,K and energy resolution 35\,meV. All samples were cleaved \textit{in-situ} and measured along the (001) surface. XAS measurements were performed using the four-circle UHV diffractometer at the REIXS 10ID-2 beamline at the Canadian Light Source in Saskatoon \cite{hawthorn2011vacuum}, with base pressure and temperature of 5\,$\cdot 10^{-10}$\,Torr and 22\,K, respectively.

\section*{Data availability}
The authors declare that the main data supporting the findings of this study are available within the paper and its Supplementary Information files. Source ARPES and XAS waves used in this study have been deposited in the Zenodo database under the digital object identifier https://doi.org/10.5281/zenodo.12759092. Additional data are available from the corresponding authors upon request.

\begin{acknowledgments}
We thank E. da Silva Neto, R. P. Day, A.M. Hallas, N. Harrison, H.-H. Kung, E. Razzoli, H. L. Tjeng, C. M. Varma, and B. Zwartsenberg for fruitful discussions. This research was undertaken thanks in part to funding from the Max Planck-UBC-UTokyo Centre for Quantum Materials and the Canada First Research Excellence Fund, Quantum Materials and Future Technologies Program. This project is also funded by the Killam, Alfred P. Sloan, and Natural Sciences and Engineering Research Council of Canada’s (NSERC’s) Steacie Memorial Fellowships (A.D.); the Alexander von Humboldt Fellowship (A.D.); the Canada Research Chairs Program (A.D.); NSERC, Canada Foundation for Innovation (CFI); the Department of National Defence (DND); British Columbia Knowledge Development Fund (BCKDF); and the CIFAR Quantum Materials Program. Part of the research described in this work was performed at the Canadian Light Source, a national research facility of the University of Saskatchewan, which is supported by CFI, NSERC, the National Research Council (NRC), the Canadian Institutes of Health Research (CIHR), the Government of Saskatchewan, and the University of Saskatchewan. The research carried out in Aarhus was supported by the Independent Research Fund Denmark (Grant No. 1026-00089B) and the VILLUM FONDEN via the Centre of Excellence for Dirac Materials (Grant No. 11744). Work at Los Alamos was performed under the auspices of the U.S. Department of Energy, Office of Basic Energy Sciences, Division of Materials Science and Engineering.
\end{acknowledgments}

\vspace{0mm}
\section*{Author contributions}
M.Z., I.S.E, G.A.S and A.D. conceived the project and M.Z. and A.D. designed the experiment. P.F.S.R. and Z.F. grew the single crystals and P.F.S.R. characterized them. M.Z. and M.M. performed the ARPES experiments, with assistance from F.B., G.L., K.V, D.C, M.B and Ph.H.. XAS measurements were carried out by M.Z and R.J.G.. M.Z. analyzed the ARPES data with input from M.M., F.B. and A.D.; R.J.G. analyzed the XAS data; M.Z. and R.J.G developed the phenomenological model. M.Z. and A.D. wrote the manuscript with input from all authors. A.D. was responsible for the overall direction, planning, and management of the project.

\section*{Competing Interests }
The authors declare no competing interests.


\begin{thebibliography}{10}
\expandafter\ifx\csname url\endcsname\relax
  \def\url#1{\texttt{#1}}\fi
\expandafter\ifx\csname urlprefix\endcsname\relax\def\urlprefix{URL }\fi
\providecommand{\bibinfo}[2]{#2}
\providecommand{\eprint}[2][]{\url{#2}}

\bibitem{mathur1998magnetically}
\bibinfo{author}{Mathur, N.~D.} \emph{et~al.}
\newblock \bibinfo{title}{Magnetically mediated superconductivity in heavy fermion compounds}.
\newblock \emph{\bibinfo{journal}{Nature}} \textbf{\bibinfo{volume}{394}}, \bibinfo{pages}{39--43} (\bibinfo{year}{1998}).

\bibitem{moriya2003antiferromagnetic}
\bibinfo{author}{Moriya, T.} \& \bibinfo{author}{Ueda, K.}
\newblock \bibinfo{title}{Antiferromagnetic spin fluctuation and superconductivity}.
\newblock \emph{\bibinfo{journal}{Reports on Progress in Physics}} \textbf{\bibinfo{volume}{66}}, \bibinfo{pages}{1299} (\bibinfo{year}{2003}).

\bibitem{hotta2006orbital}
\bibinfo{author}{Hotta, T.}
\newblock \bibinfo{title}{Orbital ordering phenomena in d-and f-electron systems}.
\newblock \emph{\bibinfo{journal}{Reports on Progress in Physics}} \textbf{\bibinfo{volume}{69}}, \bibinfo{pages}{2061} (\bibinfo{year}{2006}).

\bibitem{gegenwart2008quantum}
\bibinfo{author}{Gegenwart, P.}, \bibinfo{author}{Si, Q.} \& \bibinfo{author}{Steglich, F.}
\newblock \bibinfo{title}{Quantum criticality in heavy-fermion metals}.
\newblock \emph{\bibinfo{journal}{Nature Physics}} \textbf{\bibinfo{volume}{4}}, \bibinfo{pages}{186--197} (\bibinfo{year}{2008}).

\bibitem{fisk1988heavy}
\bibinfo{author}{Fisk, Z.} \emph{et~al.}
\newblock \bibinfo{title}{Heavy-electron metals: New highly correlated states of matter}.
\newblock \emph{\bibinfo{journal}{Science}} \textbf{\bibinfo{volume}{239}}, \bibinfo{pages}{33--42} (\bibinfo{year}{1988}).

\bibitem{coleman2015heavy}
\bibinfo{author}{Coleman, P.}
\newblock \bibinfo{title}{Heavy fermions and the kondo lattice: a 21st century perspective}.
\newblock \emph{\bibinfo{journal}{arXiv preprint arXiv:1509.05769}}  (\bibinfo{year}{2015}).

\bibitem{kondo1964resistance}
\bibinfo{author}{Kondo, J.}
\newblock \bibinfo{title}{Resistance minimum in dilute magnetic alloys}.
\newblock \emph{\bibinfo{journal}{Progress of Theoretical Physics}} \textbf{\bibinfo{volume}{32}}, \bibinfo{pages}{37--49} (\bibinfo{year}{1964}).

\bibitem{varma1976mixed}
\bibinfo{author}{Varma, C.~M.}
\newblock \bibinfo{title}{Mixed-valence compounds}.
\newblock \emph{\bibinfo{journal}{Reviews of Modern Physics}} \textbf{\bibinfo{volume}{48}}, \bibinfo{pages}{219} (\bibinfo{year}{1976}).

\bibitem{peter2007molecules}
\bibinfo{author}{Day, P.}
\newblock \emph{\bibinfo{title}{Molecules Into Materials: Case Studies in Materials Chemistry-Mixed Valency, Magnetism and Superconductivity}} (\bibinfo{publisher}{World Scientific}, \bibinfo{year}{2007}).

\bibitem{lawrence1981valence}
\bibinfo{author}{Lawrence, J.~M.}, \bibinfo{author}{Riseborough, P.~S.} \& \bibinfo{author}{Parks, R.~D.}
\newblock \bibinfo{title}{Valence fluctuation phenomena}.
\newblock \emph{\bibinfo{journal}{Reports on Progress in Physics}} \textbf{\bibinfo{volume}{44}}, \bibinfo{pages}{1} (\bibinfo{year}{1981}).

\bibitem{buschow1979intermetallic}
\bibinfo{author}{Buschow, K.}
\newblock \bibinfo{title}{Intermetallic compounds of rare earths and non-magnetic metals}.
\newblock \emph{\bibinfo{journal}{Reports on Progress in Physics}} \textbf{\bibinfo{volume}{42}}, \bibinfo{pages}{1373} (\bibinfo{year}{1979}).

\bibitem{coey1999mixed}
\bibinfo{author}{Coey, J. M.~D.}, \bibinfo{author}{Viret, M.} \& \bibinfo{author}{Von~Molnar, S.}
\newblock \bibinfo{title}{Mixed-valence manganites}.
\newblock \emph{\bibinfo{journal}{Advances in Physics}} \textbf{\bibinfo{volume}{48}}, \bibinfo{pages}{167--293} (\bibinfo{year}{1999}).

\bibitem{parks2012valence}
\bibinfo{author}{Parks, R.}
\newblock \emph{\bibinfo{title}{Valence instabilities and related narrow-band phenomena}} (\bibinfo{publisher}{Springer Science \& Business Media}, \bibinfo{year}{2012}).

\bibitem{Anderson1984}
\bibinfo{author}{Anderson, P.~W.}
\newblock \emph{\bibinfo{title}{Present Status of Theory: 1/{N} Approach}}, \bibinfo{pages}{313--326} (\bibinfo{publisher}{Springer US}, \bibinfo{year}{1984}).

\bibitem{Riseborough2016}
\bibinfo{author}{Riseborough, P.~S.} \& \bibinfo{author}{Lawrence, J.~M.}
\newblock \bibinfo{title}{Mixed valent metals}.
\newblock \emph{\bibinfo{journal}{Reports on Progress in Physics}} \textbf{\bibinfo{volume}{79}}, \bibinfo{pages}{084501} (\bibinfo{year}{2016}).

\bibitem{Riseborough2000}
\bibinfo{author}{Riseborough, P.~S.}
\newblock \bibinfo{title}{Heavy fermion semiconductors}.
\newblock \emph{\bibinfo{journal}{Advances in Physics}} \textbf{\bibinfo{volume}{49}}, \bibinfo{pages}{257--320} (\bibinfo{year}{2000}).

\bibitem{BickersRMP}
\bibinfo{author}{Bickers, N.~E.}
\newblock \bibinfo{title}{Review of techniques in the large-$n$ expansion for dilute magnetic alloys}.
\newblock \emph{\bibinfo{journal}{Rev. Mod. Phys.}} \textbf{\bibinfo{volume}{59}}, \bibinfo{pages}{845--939} (\bibinfo{year}{1987}).

\bibitem{QimiaoSi1996}
\bibinfo{author}{Si, Q.}
\newblock \bibinfo{title}{Non-fermi liquids in the extended hubbard model}.
\newblock \emph{\bibinfo{journal}{Journal of Physics: Condensed Matter}} \textbf{\bibinfo{volume}{8}}, \bibinfo{pages}{9953} (\bibinfo{year}{1996}).

\bibitem{haldane1977new}
\bibinfo{author}{Haldane, F. D.~M.}
\newblock \bibinfo{title}{New model for the mixed-valence phenomenon in rare-earth materials}.
\newblock \emph{\bibinfo{journal}{Physical Review B}} \textbf{\bibinfo{volume}{15}}, \bibinfo{pages}{2477} (\bibinfo{year}{1977}).

\bibitem{VarmaHeinePRB}
\bibinfo{author}{Varma, C.~M.} \& \bibinfo{author}{Heine, V.}
\newblock \bibinfo{title}{Valence transitions in rare-earth chalcogenides}.
\newblock \emph{\bibinfo{journal}{Phys. Rev. B}} \textbf{\bibinfo{volume}{11}}, \bibinfo{pages}{4763--4767} (\bibinfo{year}{1975}).

\bibitem{perakis1993model}
\bibinfo{author}{Perakis, I.~E.}, \bibinfo{author}{Varma, C.~M.} \& \bibinfo{author}{Ruckenstein, A.~E.}
\newblock \bibinfo{title}{Non-fermi-liquid states of a magnetic ion in a metal}.
\newblock \emph{\bibinfo{journal}{Phys. Rev. Lett.}} \textbf{\bibinfo{volume}{70}}, \bibinfo{pages}{3467--3470} (\bibinfo{year}{1993}).

\bibitem{sire1994model}
\bibinfo{author}{Sire, C.}, \bibinfo{author}{Varma, C.~M.}, \bibinfo{author}{Ruckenstein, A.~E.} \& \bibinfo{author}{Giamarchi, T.}
\newblock \bibinfo{title}{Theory of the marginal-{F}ermi-liquid spectrum and pairing in a local copper oxide model}.
\newblock \emph{\bibinfo{journal}{Phys. Rev. Lett.}} \textbf{\bibinfo{volume}{72}}, \bibinfo{pages}{2478--2481} (\bibinfo{year}{1994}).

\bibitem{tarascon1980temperature}
\bibinfo{author}{Tarascon, J.-M.}, \bibinfo{author}{Isikawa, Y.}, \bibinfo{author}{Chevalier, J., B.and~Etourneau}, \bibinfo{author}{Hagenmuller, P.} \& \bibinfo{author}{Kasaya, M.}
\newblock \bibinfo{title}{Temperature dependence of the samarium oxidation state in $\mathrm{SmB}_6$ and $\mathrm{Sm}_{1-x}\mathrm{La}_x\mathrm{B}_6$}.
\newblock \emph{\bibinfo{journal}{Journal de Physique}} \textbf{\bibinfo{volume}{41}}, \bibinfo{pages}{1141--1145} (\bibinfo{year}{1980}).

\bibitem{mizumaki2009temperature}
\bibinfo{author}{Mizumaki, M.}, \bibinfo{author}{Tsutsui, S.} \& \bibinfo{author}{Iga, F.}
\newblock \bibinfo{title}{Temperature dependence of {S}m valence in $\mathrm{SmB}_6$ studied by {X}-ray absorption spectroscopy}.
\newblock In \emph{\bibinfo{booktitle}{Journal of Physics: Conference Series}}, vol. \bibinfo{volume}{176}, \bibinfo{pages}{012034} (\bibinfo{organization}{IOP Publishing}, \bibinfo{year}{2009}).

\bibitem{butch2016pressure}
\bibinfo{author}{Butch, N.~P.} \emph{et~al.}
\newblock \bibinfo{title}{Pressure-resistant intermediate valence in the {K}ondo insulator $\mathrm{SmB}_6$}.
\newblock \emph{\bibinfo{journal}{Physical Review Letters}} \textbf{\bibinfo{volume}{116}}, \bibinfo{pages}{156401} (\bibinfo{year}{2016}).

\bibitem{sun2016pressure}
\bibinfo{author}{Sun, L.} \& \bibinfo{author}{Wu, Q.}
\newblock \bibinfo{title}{Pressure-induced exotic states in rare earth hexaborides}.
\newblock \emph{\bibinfo{journal}{Reports on Progress in Physics}} \textbf{\bibinfo{volume}{79}}, \bibinfo{pages}{084503} (\bibinfo{year}{2016}).

\bibitem{lee2017fluctuating}
\bibinfo{author}{Lee, J.-M.} \emph{et~al.}
\newblock \bibinfo{title}{The fluctuating population of $\mathrm{Sm}$ 4f configurations in topological {K}ondo insulator $\mathrm{SmB}_6$ explored with high-resolution {X}-ray absorption and emission spectra}.
\newblock \emph{\bibinfo{journal}{Dalton Transactions}} \textbf{\bibinfo{volume}{46}}, \bibinfo{pages}{11664--11668} (\bibinfo{year}{2017}).

\bibitem{Zhou2017pressure}
\bibinfo{author}{Zhou, Y.} \emph{et~al.}
\newblock \bibinfo{title}{Quantum phase transition and destruction of {K}ondo effect in pressurized $\mathrm{SmB}_6$}.
\newblock \emph{\bibinfo{journal}{Science Bulletin}} \textbf{\bibinfo{volume}{62}}, \bibinfo{pages}{1439--1444} (\bibinfo{year}{2017}).

\bibitem{emi2018temperature}
\bibinfo{author}{Emi, N.} \emph{et~al.}
\newblock \bibinfo{title}{Temperature and pressure dependences of {S}m valence in intermediate valence compound $\mathrm{SmB}_6$}.
\newblock \emph{\bibinfo{journal}{Physica B: Condensed Matter}} \textbf{\bibinfo{volume}{536}}, \bibinfo{pages}{197--199} (\bibinfo{year}{2018}).

\bibitem{menth1969magnetic}
\bibinfo{author}{Menth, A.}, \bibinfo{author}{Buehler, E.} \& \bibinfo{author}{Geballe, T.~H.}
\newblock \bibinfo{title}{Magnetic and semiconducting properties of $\mathrm{SmB}_6$}.
\newblock \emph{\bibinfo{journal}{Physical Review Letters}} \textbf{\bibinfo{volume}{22}}, \bibinfo{pages}{295} (\bibinfo{year}{1969}).

\bibitem{dzero2010topological}
\bibinfo{author}{Dzero, M.}, \bibinfo{author}{Sun, K.}, \bibinfo{author}{Galitski, V.} \& \bibinfo{author}{Coleman, P.}
\newblock \bibinfo{title}{Topological {Ko}ndo insulators}.
\newblock \emph{\bibinfo{journal}{Physical Review Letters}} \textbf{\bibinfo{volume}{104}}, \bibinfo{pages}{106408} (\bibinfo{year}{2010}).

\bibitem{dzero2012theory}
\bibinfo{author}{Dzero, M.}, \bibinfo{author}{Sun, K.}, \bibinfo{author}{Coleman, P.} \& \bibinfo{author}{Galitski, V.}
\newblock \bibinfo{title}{Theory of topological {K}ondo insulators}.
\newblock \emph{\bibinfo{journal}{Physical Review B}} \textbf{\bibinfo{volume}{85}}, \bibinfo{pages}{045130} (\bibinfo{year}{2012}).

\bibitem{wolgast2013low}
\bibinfo{author}{Wolgast, S.} \emph{et~al.}
\newblock \bibinfo{title}{Low-temperature surface conduction in the {K}ondo insulator $\mathrm{SmB}_6$}.
\newblock \emph{\bibinfo{journal}{Physical Review B}} \textbf{\bibinfo{volume}{88}}, \bibinfo{pages}{180405} (\bibinfo{year}{2013}).

\bibitem{zhang2013hybridization}
\bibinfo{author}{Zhang, X.} \emph{et~al.}
\newblock \bibinfo{title}{Hybridization, inter-ion correlation, and surface states in the {K}ondo insulator $\mathrm{SmB}_6$}.
\newblock \emph{\bibinfo{journal}{Physical Review X}} \textbf{\bibinfo{volume}{3}}, \bibinfo{pages}{011011} (\bibinfo{year}{2013}).

\bibitem{kim2013surface}
\bibinfo{author}{Kim, D.~J.} \emph{et~al.}
\newblock \bibinfo{title}{Surface {H}all effect and nonlocal transport in $\mathrm{SmB}_6$: evidence for surface conduction}.
\newblock \emph{\bibinfo{journal}{Scientific Reports}} \textbf{\bibinfo{volume}{3}}, \bibinfo{pages}{3150} (\bibinfo{year}{2013}).

\bibitem{jiang2013observation}
\bibinfo{author}{Jiang, J.} \emph{et~al.}
\newblock \bibinfo{title}{Observation of possible topological in-gap surface states in the {K}ondo insulator $\mathrm{SmB}_6$ by photoemission}.
\newblock \emph{\bibinfo{journal}{Nature Communications}} \textbf{\bibinfo{volume}{4}}, \bibinfo{pages}{3010} (\bibinfo{year}{2013}).

\bibitem{xu2013surface}
\bibinfo{author}{Xu, N.} \emph{et~al.}
\newblock \bibinfo{title}{Surface and bulk electronic structure of the strongly correlated system $\mathrm{SmB}_6$ and implications for a topological {K}ondo insulator}.
\newblock \emph{\bibinfo{journal}{Physical Review B}} \textbf{\bibinfo{volume}{88}}, \bibinfo{pages}{121102} (\bibinfo{year}{2013}).

\bibitem{neupane2013surface}
\bibinfo{author}{Neupane, M.} \emph{et~al.}
\newblock \bibinfo{title}{Surface electronic structure of the topological {K}ondo-insulator candidate correlated electron system $\mathrm{SmB}_6$}.
\newblock \emph{\bibinfo{journal}{Nature Communications}} \textbf{\bibinfo{volume}{4}}, \bibinfo{pages}{2991} (\bibinfo{year}{2013}).

\bibitem{frantzeskakis2013kondo}
\bibinfo{author}{Frantzeskakis, E.} \emph{et~al.}
\newblock \bibinfo{title}{Kondo hybridization and the origin of metallic states at the (001) surface of $\mathrm{SmB}_6$}.
\newblock \emph{\bibinfo{journal}{Physical Review X}} \textbf{\bibinfo{volume}{3}}, \bibinfo{pages}{041024} (\bibinfo{year}{2013}).

\bibitem{zhu2013polarity}
\bibinfo{author}{Zhu, Z.-H.} \emph{et~al.}
\newblock \bibinfo{title}{Polarity-driven surface metallicity in $\mathrm{SmB}_6$}.
\newblock \emph{\bibinfo{journal}{Physical Review Letters}} \textbf{\bibinfo{volume}{111}}, \bibinfo{pages}{216402} (\bibinfo{year}{2013}).

\bibitem{hlawenka2018samarium}
\bibinfo{author}{Hlawenka, P.} \emph{et~al.}
\newblock \bibinfo{title}{Samarium hexaboride is a trivial surface conductor}.
\newblock \emph{\bibinfo{journal}{Nature Communications}} \textbf{\bibinfo{volume}{9}}, \bibinfo{pages}{517} (\bibinfo{year}{2018}).

\bibitem{varma2020majoranas}
\bibinfo{author}{Varma, C.~M.}
\newblock \bibinfo{title}{Majoranas in mixed-valence insulators}.
\newblock \emph{\bibinfo{journal}{Physical Review B}} \textbf{\bibinfo{volume}{102}}, \bibinfo{pages}{155145} (\bibinfo{year}{2020}).

\bibitem{li2014two}
\bibinfo{author}{Li, G.} \emph{et~al.}
\newblock \bibinfo{title}{Two-dimensional {F}ermi surfaces in {K}ondo insulator $\mathrm{SmB}_6$}.
\newblock \emph{\bibinfo{journal}{Science}} \textbf{\bibinfo{volume}{346}}, \bibinfo{pages}{1208--1212} (\bibinfo{year}{2014}).

\bibitem{tan2015unconventional}
\bibinfo{author}{Tan, B.~S.} \emph{et~al.}
\newblock \bibinfo{title}{Unconventional {F}ermi surface in an insulating state}.
\newblock \emph{\bibinfo{journal}{Science}} \textbf{\bibinfo{volume}{349}}, \bibinfo{pages}{287--290} (\bibinfo{year}{2015}).

\bibitem{Thomas2019}
\bibinfo{author}{Thomas, S.~M.} \emph{et~al.}
\newblock \bibinfo{title}{Quantum oscillations in flux-grown $\mathrm{SmB}_6$ with embedded aluminum}.
\newblock \emph{\bibinfo{journal}{Phys. Rev. Lett.}} \textbf{\bibinfo{volume}{122}}, \bibinfo{pages}{166401} (\bibinfo{year}{2019}).

\bibitem{sundermann4fcrystalfield}
\bibinfo{author}{Sundermann, M.} \emph{et~al.}
\newblock \bibinfo{title}{4$f$ crystal field ground state of the strongly correlated topological insulator $\mathrm{SmB}_6$}.
\newblock \emph{\bibinfo{journal}{Physical Review Letters}} \textbf{\bibinfo{volume}{120}}, \bibinfo{pages}{016402} (\bibinfo{year}{2018}).

\bibitem{Goodrich2009dHvA}
\bibinfo{author}{Goodrich, R.~G.}, \bibinfo{author}{Young, D.~P.}, \bibinfo{author}{Harrison, N.}, \bibinfo{author}{Capan, C.} \& \bibinfo{author}{Fisk, Z.}
\newblock \bibinfo{title}{Fermi surfaces changes in La$_{1-x}$Sm$_x$B$_6$ and Ce$_{1-x}$Ca$_x$B$_6$ studied using the de Haas--van Alphen effect and magnetic susceptibility}.
\newblock \emph{\bibinfo{journal}{Phys. Rev. B}} \textbf{\bibinfo{volume}{80}}, \bibinfo{pages}{233101} (\bibinfo{year}{2009}).
\newblock \bibinfo{note}{Please note that for a quantitative comparison with our work, the raw FS volume values presented in this reference should be used and not the percentage volume reductions which were multiplied by an additional factor of 3.}

\bibitem{denlinger2014smb6}
\bibinfo{author}{Denlinger, J.~D.} \emph{et~al.}
\newblock \bibinfo{title}{$\mathrm{SmB}_6$ photoemission: past and present}.
\newblock In \emph{\bibinfo{booktitle}{Proceedings of the International Conference on Strongly Correlated Electron Systems (SCES2013)}}, \bibinfo{pages}{017038} (\bibinfo{year}{2014}).

\bibitem{martin1982fermi}
\bibinfo{author}{Martin, R.~M.}
\newblock \bibinfo{title}{Fermi-surface sum rule and its consequences for periodic kondo and mixed-valence systems}.
\newblock \emph{\bibinfo{journal}{Physical Review Letters}} \textbf{\bibinfo{volume}{48}}, \bibinfo{pages}{362} (\bibinfo{year}{1982}).

\bibitem{martin1982fermi2}
\bibinfo{author}{Martin, R.~M.}
\newblock \bibinfo{title}{The fermi surface and {F}ermi liquid properties of periodic {K}ondo and mixed valence systems}.
\newblock \emph{\bibinfo{journal}{Journal of Applied Physics}} \textbf{\bibinfo{volume}{53}}, \bibinfo{pages}{2134--2136} (\bibinfo{year}{1982}).

\bibitem{kasaya1980study}
\bibinfo{author}{Kasaya, M.}, \bibinfo{author}{Tarascon, J.~M.} \& \bibinfo{author}{Etourneau, J.}
\newblock \bibinfo{title}{Study of the valence transition in {L}a- and {Y}b-substituted $\mathrm{SmB}_6$}.
\newblock \emph{\bibinfo{journal}{Solid State Communications}} \textbf{\bibinfo{volume}{33}}, \bibinfo{pages}{1005--1007} (\bibinfo{year}{1980}).

\bibitem{tarascon1980valence}
\bibinfo{author}{Tarascon, J.-M.} \emph{et~al.}
\newblock \bibinfo{title}{Valence transition of samarium in hexaboride solid solutions {S}m$_{1-x}${M}$_x${B}$_6$ ({M}= {Y}b$^{2+}$, {S}r$^{2+}$, {L}a$^{3+}$, {Y}$^{3+}$, {T}h$^{4+}$)}.
\newblock \emph{\bibinfo{journal}{Journal de Physique}} \textbf{\bibinfo{volume}{41}}, \bibinfo{pages}{1135--1140} (\bibinfo{year}{1980}).

\bibitem{kang2016magnetic}
\bibinfo{author}{Kang, B.~Y.} \emph{et~al.}
\newblock \bibinfo{title}{Magnetic and nonmagnetic doping dependence of the conducting surface states in $\mathrm{SmB}_6$}.
\newblock \emph{\bibinfo{journal}{Physical Review B}} \textbf{\bibinfo{volume}{94}}, \bibinfo{pages}{165102} (\bibinfo{year}{2016}).

\bibitem{gabani2016transport}
\bibinfo{author}{Gab{\'a}ni, S.} \emph{et~al.}
\newblock \bibinfo{title}{Transport properties of variously doped $\mathrm{SmB}_6$}.
\newblock \emph{\bibinfo{journal}{Philosophical Magazine}} \textbf{\bibinfo{volume}{96}}, \bibinfo{pages}{3274--3283} (\bibinfo{year}{2016}).

\bibitem{hoffmann2004undulator}
\bibinfo{author}{Hoffmann, S.~V.}, \bibinfo{author}{S{\o}ndergaard, C.}, \bibinfo{author}{Schultz, C.}, \bibinfo{author}{Li, Z.} \& \bibinfo{author}{Hofmann, P.}
\newblock \bibinfo{title}{An undulator-based spherical grating monochromator beamline for angle-resolved photoemission spectroscopy}.
\newblock \emph{\bibinfo{journal}{Nuclear Instruments and Methods in Physics Research Section A: Accelerators, Spectrometers, Detectors and Associated Equipment}} \textbf{\bibinfo{volume}{523}}, \bibinfo{pages}{441--453} (\bibinfo{year}{2004}).

\bibitem{hawthorn2011vacuum}
\bibinfo{author}{Hawthorn, D.~G.} \emph{et~al.}
\newblock \bibinfo{title}{An in-vacuum diffractometer for resonant elastic soft x-ray scattering}.
\newblock \emph{\bibinfo{journal}{Review of Scientific Instruments}} \textbf{\bibinfo{volume}{82}}, \bibinfo{pages}{073104} (\bibinfo{year}{2011}).

\end{thebibliography}
\end{document}